\documentstyle[12pt]{article}
\newcommand{\cd}{c^{\dagger}}
\newcommand{\Tr}{{\rm Tr}}
\newcommand{\abs}[1]{\left|#1\right|}
\newcommand{\bkt}[1]{\left\langle#1\right\rangle}
\newcommand{\rbk}[1]{\left(#1\right)}
\newcommand{\sqbk}[1]{\left[#1\right]}
\renewcommand{\phi}{\varphi}
\newcommand{\ret}{\nonumber\\}
\newcommand{\normi}[1]{\Vert#1\Vert_{\infty}}
\newcommand{\up}{\uparrow}
\newcommand{\dn}{\downarrow}
\begin{document}
\begin{center}
{\Large\bf Decay of Superconducting and Magnetic
Correlations in One- and Two-Dimensional
Hubbard Models\footnote{
Published in Phys. Rev. Lett. {\bf 68}, 2348 (1992).
Reference \cite{MacrisRuiz} (which appeared after the publication) was added in
the archived version.
}}
\par\bigskip
{Tohru Koma\footnote{
koma@riron.gakushuin.ac.jp
} and Hal Tasaki\footnote{
hal.tasaki@gakushuin.ac.jp,
http://www.gakushuin.ac.jp/$\tilde{\ }$881791/halE.htm
}}
\par\bigskip
{\small\em Department of Physics, Gakushuin University,
Mejiro, Toshima-ku, Tokyo 171, JAPAN}
\end{center}
\begin{abstract}
In a general class of one and two dimensional Hubbard models, 
we prove upper bounds for the two-point correlation functions 
at finite temperatures for electrons, for electron pairs, 
and for spins.  
The upper bounds decay exponentially in one dimension, 
and with power laws in two dimensions.
The bounds rule out the possibility of the corresponding 
condensation of superconducting electron pairs, 
and of the corresponding magnetic ordering.  
Our method is general enough to cover 
other models such as the \( t \)-\( J \) model.
\end{abstract}

The Hubbard model and its variants have been 
attracting considerable interest.  
But rigorous results are still rare.  
In one dimension, the Bethe ansatz method has been 
successfully applied \cite{1} both to the ground state 
and to the finite temperature Gibbs state.  
In general dimensions, Lieb's theorem \cite{2} and 
Nagaoka's theorem \cite{3} on the ground state 
structures are known.  In one and two dimensions, 
Ghosh \cite{4} proved the absence of 
magnetic ordering at finite temperatures.

In the present letter, we extend McBryan and Spencer's 
method \cite{5} developed in 
classical spin systems to a general class of 
Hubbard models in one and two dimensions, 
and prove upper bounds for various correlation 
functions at finite temperatures.
The 
bounds rule out the possibility of magnetic 
ordering and condensation of electrons or 
superconducting electron pairs such as Cooper 
pairs or \( \eta \)-pairs \cite{6}.

We consider a tight binding electron model on the one 
dimensional lattice \( \bf Z \) or the 
square lattice \( {\bf Z}^{2} \) \cite{7}.  
The Hamiltonian is given by
\begin{equation}
	H = -\sum_{x,y\in{\bf Z}^{d}}\sum_{\sigma=\up,\dn}
	t_{x,y}c^{\dagger}_{x,\sigma}c_{y,\sigma}
	+V(\{n_{x,\sigma}\})
	+\sum_{x\in{\bf Z}^{d}}{\bf h}_{x}\cdot{\bf S}_{x},
	\label{1}
\end{equation}
with \( d=1 \) or \( 2 \). 
The number operators are defined by
\( n_{x,\sigma}=c^{\dagger}_{x,\sigma}c_{x,\sigma} \),
and spin operators by 
\( S^{j}_{x}=\sum_{\sigma,\sigma'=\up,\dn}
c^{\dagger}_{x,\sigma}\tau^{j}_{\sigma,\sigma'}c_{x,\sigma'} \)
(\( j = 1,2,3 \)),
where 
\( (\tau^{j}_{\sigma,\sigma'})_{\sigma,\sigma'=\up,\dn} \)
are Pauli spin matrices and 
\( c^{\dagger}_{x,\sigma} \), \( c_{x,\sigma} \),
are the creation and the anihilation operators, respectively, 
for the electron at site \( x \) with 
spin \( \sigma \).  
The hermitian hopping matrix \( ( t_{xy} ) \) is arbitrary, 
except for the conditions that 
there are finite constants \( t \), \( R \), 
and 
\( |t_{xy}| \le t  \)
holds for any  \( x \), \( y \), 
and \( t_{xy} \) is vanishing \cite{8} for 
\( | x - y | \ge R \).
Note that we can include external magnetic 
field which is represented by 
complex \( t_{xy} \).  
The interaction \( V(\{n_{x,\sigma}\}) \) 
is an arbitrary function of the number operators, 
and \( {\bf h}_{x} \) represents local magnetic field 
or spin-flip impurity.  
Note that the Hamiltonian (\ref{1}) 
is not  necessarily completely isotropic 
in spin space, but has a global \( O(2) \) symmetry 
related to the spin rotation about the \( z \)-axis.  
We stress that the class of Hamiltonians 
considered here includes not only the well 
studied models like the (standard) Hubbard 
model or the periodic Anderson model, 
but also many of their variants with , e.g., long-range, 
random or spin-dependent interactions.  

To define the Gibbs state, we replace the infinite 
lattice with a finite lattice of linear 
dimension \( L \) with periodic boundary conditions.
The thermal expectation value of an 
arbitrary operator \( A \) is defined by 
\begin{equation}
	\bkt{A}_{L}  = 
	\frac{\Tr( A e^{-\beta H} ) }{ \Tr( e^{-\beta H} )},
\end{equation}
where the trace is 
over all the electron states.
We consider the infinite volume state defined by
\begin{equation}
	\bkt{A}=\lim_{L\to\infty}\bkt{A}_{L},
\end{equation}
with the electron density fixed to \( \rho \).
Our result is independent of \( \rho \) and thus 
applies to grand canonical averages as well.

The main result of the present letter is the following.

\par\bigskip\noindent
{\bf Theorem:}  There exist finite constants \cite{9}
\( \alpha,\gamma,\delta \) and a function 
\( f(\beta) \) which depend only 
on the hopping matrix \( ( t_{xy} ) \).  
The function \( f(\beta) \) is decreasing and behaves as 
\( f(\beta) \approx 1/\beta\) for 
\( \beta\gg\delta \)
and
\( f(\beta) \approx (2/\delta)|\ln\beta| \) for 
\( \beta\ll\delta \).
In a two dimensional model in the class described 
above, we have
\begin{eqnarray}
	\abs{\bkt{
	\cd_{x,\up}\cd_{x,\dn}c_{y,\up}c_{y,\dn}+\mbox{H.c.}
	}} 
	& \le & 
	2|x-y|^{-\alpha f(\beta)},
	\label{2}  \\
	\abs{
	\bkt{
	\cd_{x,\sigma}c_{y,\sigma}+\mbox{H.c.}
	}
	} 
	& \le & 
	2|x-y|^{-\alpha f(2\beta)/2},
	\label{3}
\end{eqnarray}
for any finite \( \beta \) and for any \( x \), \( y \) 
with sufficiently large 
\( |x-y| \).  
If the local field has the form 
\( {\bf h}_{x}=(0,0,h_{x}) \)
we further have
\begin{equation}
	\abs{\bkt{
	S_{x}^{1}S_{y}^{1}+S_{x}^{2}S_{y}^{2}
	}}
	\le
	|x-y|^{-\alpha f(\beta)},
	\label{4}
\end{equation}
for any finite \( \beta \) and for any \( x, y \) 
with sufficiently large \( |x-y| \).  
In a one dimensional model, 
we have the above bounds (\ref{2}), (\ref{3}) and (\ref{4}) 
with the right-hand-sides replaced with
\( 2 \exp[ -\gamma f(\beta)|x-y| ] \), 
\( 2 \exp[ -\gamma f(2\beta) |x-y|/2 ] \) 
and 
\( \exp[ -\gamma f(\beta)|x-y| ] \), 
respectively.
\par\bigskip
The above bounds rigorously rule out the 
possibility of the corresponding 
condensations of electrons or electron pairs and of the 
corresponding magnetic ordering.  
The bound (\ref{2}), for example, inhibits the condensation 
of singlet electron pairs such as the 
Cooper pairs or the \( \eta \)-pairs \cite{6}.  
However our method can be easily extended to rule out 
any kind of condensation which is related to a 
spontaneous breakdown of the quantum 
mechanical global \( U(1) \) symmetry.  
It is also straightforward to extend the method to cover 
other systems such as the Hubbard model with
nonlocal spin-flip term or the \( t \)-\( J \) model 
\cite{10}.
The explicit upper bounds for the 
correlation functions provide further information 
about the propagation of electrons, electron pairs and magnons.  
The astonishing generality 
of the theorem, especially the complete arbitrariness 
of interactions, may be regarded as a 
sharp demonstration of the fact that the electron hopping plays 
a fundamental role in 
various condensation phenomena in itinerant electron systems.

The power law decaying upper bounds in the 
theorem are certainly not optimal at 
high temperatures, where one generally 
expects to have exponential decay.  
Even in low 
temperatures, a class of models which 
are sufficiently close to the antiferromagnetic 
Heisenberg model is expected to show exponential decay.
Among the varieties of models 
covered by the theorem, however, one might 
well find those which exhibit ``exotic'' phase 
transitions leading to power law decay.  
It is notable that the power indices in the upper 
bounds (\ref{2}), (\ref{3}), (\ref{4}) are proportional to
\( \beta^{-1} \) at low temperatures.  
This means that the 
slowest possible decay in these models is 
of the Kosterlitz-Thouless type.  
In one 
dimension, the exponentially decaying upper 
bounds in the theorem provide upper bounds 
for various correlation lengths.  
The bounds, which are proportional to \( \beta^{-1} \)  at low 
temperatures and to \( |\ln\beta| \) at high temperatures, 
reproduce a typical crossover behavior of 
correlation lengths in one dimensional 
tight-binding electron systems.

Our proof is based on the method developed 
by McBryan and Spencer \cite{5} for 
classical spin systems, and on its extension 
to quantum spin systems by Ito \cite{11}.  
In these 
works, the global continuous symmetry of the 
spin space played an essential role \cite{12}.  
Our strategy here is to make use of the global 
\( U(1) \) symmetry related to the quantum 
mechanical phase.  In this approach, 
we do not have to make further assumptions on the 
symmetry of the system since  the \( U(1) \) symmetry 
exists in any quantum particle systems.  
We believe that the present method can be 
extended to much larger class of quantum 
particle systems.  In the present letter, 
we restrict ourselves to the lattice fermion problems, 
which are free from ultraviolet divergence.

The absence of magnetic ordering in one and 
two dimensions was proved by Ghosh 
\cite{4}, who extended the Bogoliubov 
inequality method of Mermin and Wagner's \cite{13}.  
We 
note that, by combining the Mermin-Wagner 
argument with the idea to make use of the 
quantum mechanical \( U(1) \) symmetry, one can 
also prove the absence of condensation of 
electron pairs (or electrons).  
To do this, one should replace the operators 
\( A \) and \( B \) in \cite{4} 
with the Fourier transforms of the number operator 
\( n_{x}=n_{x,\up}+n_{x,\dn} \)
and of the order 
variable 
\( O_{x}=c_{x,\up}c_{x,\dn} \) (or \( c_{x,\sigma} \)), 
respectively.  
We also note that the Mermin-Wagner 
argument can be extended to cover non-translation-invariant 
models as those considered 
here.

\par\bigskip
In what follows, we describe the proof of the bound (\ref{2}) in detail.
We first prove the 
bound in a finite periodic lattice of linear dimension \( L \), 
and then take the limit \( L\to\infty \).  
To 
make use of the global quantum mechanical symmetry, 
we note that the \( U(1) \) gauge 
transformation is represented by the unitary operator
\begin{equation}
	G(\theta)=
	\prod_{u,\sigma}\exp[-i\theta_{u}n_{u,\sigma}],
	\label{5}
\end{equation}
where \( \theta=\{\theta_{u}\} \) is an arbitrary real
 function on the lattice.  
In the following, however, we 
let \( \theta_{u} \) to be pure imaginary, 
in which case the operator \( G(\theta) \) is no longer unitary.  
Since \( G(\theta) \) is invertible, we have
\begin{equation}
	\Tr[Ae^{-\beta H}]=
	\Tr\sqbk{
	G(\theta)AG(\theta)^{-1}
	\exp[-\beta G(\theta)HG(\theta)^{-1}]
	},
	\label{6}
\end{equation}
for arbitrary complex \( \theta_{u} \).  
Here the transformed Hamiltonian is
\begin{equation}
	G(\theta)HG(\theta)^{-1}
	=
	-\sum_{u,v,\sigma}
	t_{u,v}\,e^{-i(\theta_{u}-\theta_{v})}
	c^{\dagger}_{u,\sigma}c_{v,\sigma}
	+V(\{n_{u,\sigma}\})
	+\sum_{u}{\bf h}_{u}\cdot{\bf S}_{u}.
	\label{7}
\end{equation}

Let \( \phi=\{\phi_{u}\} \) be a real function 
which will be specified later.  
We consider the 
operator \( G(-i\phi) \) obtained by setting 
\( \theta=-i\phi=\{-i\phi_{u}\} \) in (\ref{5}).  
Let us fix lattice sites \( x, y \), 
and take
\( A=\cd_{x,\up}\cd_{x,\dn}c_{y,\up}c_{y,\dn} \).  
Straightforward calculations show
\begin{equation}
	G(-i\phi)AG(-i\phi)^{-1}=
	\exp[-2(\phi_{x}-\phi_{y})]A,
\end{equation}
and
\begin{equation}
	G(-i\phi)HG(-i\phi)^{-1}=
	H+U+iP,
\end{equation}
where
\begin{equation}
	U=-\sum_{u,v,\sigma}
	t_{u,v}\{\cosh(\phi_{u}-\phi_{v})-1\}
	\cd_{u,\sigma}c_{v,\sigma},
	\label{8}
\end{equation}
and
\begin{equation}
	P=-i\sum_{u,v,\sigma}
	t_{u,v}\sinh(\phi_{u}-\phi_{v})
	\cd_{u,\sigma}c_{v,\sigma}
	\label{9}
\end{equation}
are hermitian matrices.

We can bound the right-hand-side of (\ref{6}) as
\begin{eqnarray}
	&&
	\abs{\Tr\sqbk{
	G(-i\phi)AG(-i\phi)^{-1}
	\exp[-\beta G(-i\phi)HG(-i\phi)^{-1}]
	}}
	\ret&&
	\le
	e^{-2(\phi_{x}-\phi_{y})}
	(\normi{A^{*}A})^{1/2}
	\Tr[e^{-(\beta GHG^{-1})/2}e^{-(\beta G^{-1}HG)/2}]
	\ret&&
	\le
	e^{-2(\phi_{x}-\phi_{y})}
	\Tr[e^{-\beta(H+U)}]
	\ret&&
	\le
	e^{-2(\phi_{x}-\phi_{y})}
	\normi{e^{\beta U}}
	\Tr[e^{\beta H}],
	\label{10}
\end{eqnarray}
where \( \normi{O} \) denotes the maximum of the 
absolute values of the eigenvalues of a hermitian 
matrix \( O \).  
To prove the above bounds, we use the following inequalities 
for operators 
(matrices) on a finite dimensional space.  
i)~The Schwartz inequality; 
\( \Tr[OP ] \le \{\Tr[O^*O ] \Tr[ P^*P ]\}^{1/2} \) 
with \( O, P \) arbitrary.  
ii)~\( |\Tr[ O P ]| \le \normi{O} \Tr[ P ] \)
with 
\( O \) hermitian and \( P \) positive.  
iii)~\( \Tr[(O^{*})^{N} O^{N} ] \le \Tr[(O^{*}O )^{N}] \) 
with\(  N = 2^{m} \) and \( O \) arbitrary 
\cite{14}.  
iv)~The Golden-Symanzik-Thompson inequality \cite{15} 
\( \Tr[ e^{O+P} ] \le \Tr[e^{O} e^{P} ] \) where 
\( O, P \) hermitian.  
To show the first bound in (\ref{10}), we set 
\( W=\exp[-\beta G(-i\phi)HG(-i\phi)^{-1}/2] \),
\( A'=G(-i\phi)AG(-i\phi)^{-1} \), 
and use i) and ii) to get 
\begin{eqnarray}
	\Tr[(A'W)W]
	&\le&
	\rbk{
	\Tr[A'^{*}A'WW^{*}]\Tr[WW^{*}]
	}^{1/2}
	\ret
	&\le&
	(\normi{A'^{*}A'})^{1/2}\Tr[WW^{*}].
\end{eqnarray}
The second 
bound follows by noting that
\( \normi{A}=1 \), and setting 
\begin{equation}
	X=\exp\sqbk{-\frac{\beta G(-i\phi)HG(-i\phi)^{-1}}{2N}},
\end{equation}
to get 
\( \Tr[X^{N}(X^{*})^{N} ] \le \Tr[ (XX^{*})^{N} ]  \)
from iii).  
The right hand side converges to
\par\noindent 
\( \Tr[ \exp[ -\beta(H +U) ] ] \) as 
\( N\to\infty \).  
The third bound is an 
easy consequence of iv) and ii).

Now we choose \( \phi \).
Let \( \lambda_{uv} \) be real hopping matrix 
elements that satisfy 
\( \lambda_{uv}=\lambda_{vu}\ge|t_{uv}| \),
and 
\( \lambda_{uv}=0 \)
for \( |u-v| \ge R \).  
We further require \( \lambda_{uv} \) to be periodic, 
i.e., there are 
positive integers \( p, q \), and 
\( \lambda_{uv}=\lambda_{u+d,v+d} \)
holds for any 
\( d=mpe_{1}+mqe_{2} \)
where \( m, n \) 
are arbitrary integers and 
\( e_{1},e_{2} \) are two unit vectors of the lattice.  
(In one dimension, we 
of course set \( d = m p e_{1} \).)  
We assume that the lattice size \( L \) is a 
common multiple of the 
periods \( p, q \).  
The conditions imposed on \( t_{uv} \) ensures the 
existence of such  \( \lambda_{uv} \).  
(The simplest choice, which is always possible, is 
\( \lambda_{uv}=t \)
for \( |u-v| < R \), and 
\( \lambda_{uv}=0 \) otherwise.  
By choosing  \( \lambda_{uv} \) which is ``closer'' to \( t_{uv} \), 
however, one gets better constants in the 
resulting bounds.)  
Let 
\( f=\{f_{u}\} \)
be a function of the lattice sites, and define a lattice 
Laplacian \( \Delta \) by  
\( (\Delta f)_{u}=\sum_{v}\lambda_{uv}(f_{v}-f_{u}) \).
We let 
\( \phi=\{\phi_{u}\} \)
be the unique solution \cite{16} of 
the Poisson equation 
\( -(\Delta\phi)_{u}=q(\delta_{x,u}-\delta_{y,u}) \)
with a zero-point condition
\( \phi_{y}=0 \).  
The ``charge'' \( q > 0 \) will be 
determined later. 
By using the periodicity of 
\( \lambda_{uv} \)
 and explicitly writing down the solution in 
terms of the Fourier series, 
one finds that \( \phi \) has the following two 
properties \cite{5}.  
P1)~There exists a finite constant \( \delta \), and 
\( |\phi_{u}-\phi_{v}|\le q\delta \)
holds for any \( u, v \) with \( |u-v| < R \).  
P2)~In the 
\( L\to\infty \) limit, one has 
\( \phi_{x}\ge q\gamma |x-y|  \)
in one dimension and 
\( \phi_{x}\ge q\alpha \ln |x-y| \) in two 
dimensions for sufficiently large \( |x-y| \) 
with finite constants \( \gamma,\alpha \).

Noting that the above property P1) implies 
\( \cosh( \phi_{u}-\phi_{v} ) - 1 
\le g(q) ( \phi_{u}-\phi_{v} )^{2} \) 
for 
\( |u-v| < R \) 
with 
\( g(q) = \{\cosh(q\delta) - 1\} / (q\delta)^{2} \), 
we have
\begin{eqnarray}
	\normi{\exp[-\beta U]}
	&\le&
	\exp\sqbk{
	\beta\sum_{u,v}\abs{
	t_{uv}[\cosh(\phi_{u}-\phi_{v})-1]
	}}
	\ret
	&\le&
	\exp\sqbk{
	\beta g(q)
	\sum_{u,v}\lambda_{u,v}(\phi_{u}-\phi_{v})^{2}
	}
	\ret
	&=&
	\exp\sqbk{
	-2\beta g(q)\sum_{u}\phi_{u}(\Delta\phi)_{u}
	}
	\ret
	&=&
	\exp\sqbk{
	2\beta g(q)q\phi_{x}
	},
	\label{11}
\end{eqnarray}
where we have used
\( \normi{\cd_{u,\sigma}c_{v,\sigma}+
\cd_{v,\sigma}c_{u,\sigma}}=1 \)
to get the first bound.  
By substituting the 
bounds (\ref{10}) and (\ref{11}) into (\ref{6}), we get
\begin{equation}
	\abs{\bkt{A}_{L}}
	=
	\frac{\abs{\Tr[Ae^{-\beta H}]}}{\Tr[Ae^{-\beta H}]}
	\le
	\exp[-2\phi_{x}+2\beta g(q)q\phi_{x}].
	\label{12}
\end{equation}
To optimize this bound, we define
\begin{equation}
	f(\beta)
	=
	\max_{q\ge0}[2q-2\beta\delta^{-2}\{\cosh(q\delta)-1\}],
	\label{13}
\end{equation}
which is manifestly decreasing in \( \beta \), and has the 
asymptotic behavior stated in the theorem.  
By using the property P2) of \( \phi \) and letting the 
``charge'' \( q \) to be the maximizer in the above, 
we finally get
\begin{equation}
	\abs{\bkt{
	\cd_{x,\up}\cd_{x,\dn}c_{y,\up}c_{y,\dn}+\mbox{H.c.}
	}}
	\le
	\cases{
	\exp[-f(\beta)\gamma|x-y|]
	&if \( d=1 \);\cr
	\exp[-f(\beta)\alpha\ln|x-y|]=|x-y|^{-\alpha f(\beta)}
	&if \( d=2 \),
	}
	\label{14}
\end{equation}
for sufficiently large \( |x-y| \).  
Thus the bound (\ref{2}) has been proved.

The bound (\ref{3}) is proved in exactly the same manner.  
To prove the bound (\ref{4}), we set 
\( A=S_{x}^{+}S_{y}^{-} \), 
and perform a spin-dependent unitary transformation represented by 
\( G(\theta)=\prod_{u,\sigma}\exp[-i\sigma\theta_{u}n_{u,\sigma}] \).  
The rest of the proof proceeds in exactly the same way as 
the above.

\par\bigskip
We wish to thank Elliott Lieb for letting us know the reference \cite{4}.

\par\bigskip\noindent
{\em Note added (September 1997):}
In \cite{MacrisRuiz}, extensions of our results to models with long range hopping
was discussed.




\begin{thebibliography}{99}
\bibitem{8}
By using the technique developed in \cite{11}, 
it is possible to 
treat long ranged \( t_{xy} \).  
In this 
case, however, the upper bounds decay much slower than power laws.
\bibitem{9}
In the translation invariant model with 
uniform nearest neighbor hopping \( t \), we have 
\( \gamma=\delta=1/t \) in one dimension, and 
\( \alpha=(2\pi t)^{-1} \) and 
\( \delta=4/t \) in two dimensions.
\bibitem{14}
We prove the inequality for a positive integer \( m \) 
by an argument similar to that of 
Golden \cite{15}.  
Let
\( \alpha_{n}=O^{2^{m-n}}(O^{*})^{2^{m-n}} \)
and
\( \beta_{n}=(O^{*})^{2^{m-n}}O^{2^{m-n}} \)
which are hermitian.  
Note that we have 
\( \Tr[(\alpha_{n})^{N}]=\Tr[(\beta_{n})^{N}]
=\Tr[(\alpha_{n+1}\beta_{n+1})^{N}] \).
The inequality to be 
proved is
\( \Tr[\alpha_{0}]\le\Tr[(\alpha_{m})^{2^{m}}] \).  
By using i) and 
\( \Tr[(AB)^{2^{k}}]\le\Tr[A^{2^{k}}B^{2^{k}}] \)
with \( A, B \) hermitian [15], we find for \( 0 \le k \le m-1 \) 
that 
\begin{eqnarray}
	\Tr[(\alpha_{k})^{2^{k}}] 
	& = & 
	\Tr[(\alpha_{k+1}\beta_{k+1})^{2^{k}}]
	\ret
	&\le&
	\Tr[(\alpha_{k+1})^{2^{k}}(\beta_{k+1})^{2^{k}}]
	\ret
	&\le&
	\Tr[(\alpha_{k+1})^{2^{k+1}}],
\end{eqnarray}
which proves 
the desired bound.
\bibitem{16}
The Laplacian \( \Delta \) 
is not invertible since it has a zero mode 
\( f_{u}=\mbox{const} \).  However the 
Perron-Frobenius theorem ensures that the null 
space is one dimensional, and hence 
the Poisson equation with vanishing total 
charge has a unique solution.
\end{thebibliography}
\end{document}